# PHOTONIC FLAME EFFECT

N.V.Tcherniega, A.D.Kudryavtseva
P.N.Lebedev Physical Institute, RAS
Leninskii pr., 53, 119991, Moscow, Russia
tchera@mail1.lebedev.ru

## Abstract

We observed new effect which we called photonic flame effect (PFE). Several 3-dimensional photonic crystals (artificial opals) were posed on Cu plate at the temperature of liquid nitrogen (77K). Typical distance between them was 1-5 centimeters. Long-continued optical luminescence was excited in one of them by the ruby laser pulse. Analogous visible luminescence manifesting time delay appeared in other samples of the crystals. Experiments were realized for opal crystals and for nanocomposites (opals filled with nonlinear liquids).

## 1. Introduction

Photonic crystals (materials with special periodicity of dielectric constant with a period of the order of the electromagnetic wavelength) have attracted much attention since the first articles concerning such structures [1,2,3]. One-, two- and three-dimensional photonic crystals exhibit the remarkable properties which can be effectively used for photon fluxes processing due to the existing of the photonic band gap. Different types of modes defined by the periodical structure of photonic crystal and the possibility of different photonic crystal structure production lead to the important possible applications. The study of the linear optical properties of the photonic band gap have been the task of many theoretical and experimental works and still remain the task to be investigated [4,5]. The description of the electromagnetic field inside the photonic crystal structures (obtained by transfer matrix method [6] or coupled mode theory [7]) gives the clear picture of the transmitted and reflected spectrum, electromagnetic field distribution inside the crystal and their dependence on the parameters of the photonic crystal structure (values of period, number of periods, refractive index contrast). Large values of the electromagnetic field localization in some regions lead to the expectation of the strong enhancement of nonlinear wave-matter interaction in comparison with bulk crystals. Second harmonic generation in different types of photonic crystals was investigated in [8,9]. Properly chosen photonic crystal exhibits negative refraction at some conditions [10]. Some features of the stimulated Raman scattering in one-dimensional photonic structure were considered in [11]. Fully quantum mechanical treatment of the generation of entangled photon in nonlinear photonic crystals at the process of down-conversion was realized in [12]. Photonic band gap properties which are demonstrated by photonic crystalls are being actively used for investigation of photon-exciton interaction [13]. Acoustic modes excited in $SiO_2$ balls which compose opal photonic crystal show the effect of phonon modes quantization [14] and are the reason of stimulated globular scattering [15]. Specific features of the acoustic wave propagation in the photonic structures lead to the possibility of the diverging ultrasonic beam focusing into a narrow focal spot with a large focal depth [16].

The aim of this work is to study collective behavior of several photonic crystals. The crystals are posed on Cu plate at the temperature of liquid nitrogen. One of the photonic crystals is illuminated by laser pulse and the laser light is focused on this only crystal. Below we communicate observation of an effect which we call "photonic flame effect". The phenomenon which we observe is the appearance of luminescence of other photonic crystals. The duration of the luminescence of other crystals which are spatially separated with the crystal illuminated by laser pulse is of the order of seconds. The appearance of the luminescence takes place with some time delay respectively to the laser pulse. The form of these light spots on the other crystals and their slow motions along the crystal reminds a small flame spot. This inspired us to give the "photonic flame" name to the observed effect. The paper is organized as follows. In Sec.2 the experimental setup, laser, the photonic crystals (artificial opals) used in the experiment are described. In Sec.3 the "photonic flame effect" observed in the experiment is discussed. In Sec.4 perspectives and possible explanations are presented.

## 2. Photonic crystals and laser used in experiment.

One of the most promising three-dimensional photonic crystals is artificial opal. Opal is a crystal with face-centered cubic lattice consisting of the monodisperse close packed $SiO_2$ spheres with diameter about several hundred nanometers. Because the refractive index contrast (ratio $n_{SiO2}/n_{air}$) is about 1,45 the complete photonic band gap does not exist but the photonic pseudogap takes place. Empty cavities among these globules have octahedral and tetrahedral form. It is possible to investigate both initial opals (opal matrices) and nanocomposites, in which cavities are filled with organic or inorganic materials, for instance, semiconductors, superconductors, ferromagnetic substances, dielectrics, displaying different types of nonlinearities and so on. Filling voids of the photonic crystal with materials with different refraction index one can effectively process the parameters of the photonic pseudogap.
Ruby laser giant pulse ($\lambda$=694.3 nm, $\tau$=20 ns, $E_{max}$ =0.3 J, spectral width of the initial light - 0.015 $cm^{-1}$.) has been used as a source of excitation. Exciting light has been focused on the material by lenses with different focal lengths (50, 90, and 150 mm). The samples of opal crystals used had the size 3x5x5mm and were cut parallel to the plane (111) (see Fig.2) .The angle of the incidence of the laser beam on the plane (111) varied from 0 to $60^0$. Sample distance from focusing system and exciting light energy were different in different runs of the experiment. This gave possibility to make measurements for different power density at the entrance of the sample and for different field distribution inside the sample. Opal crystals consisting of the close-packed amorphose spheres with diameter 200 nm and nanocomposites (opal crystals with voids filled with acetone or ethanol) were investigated.

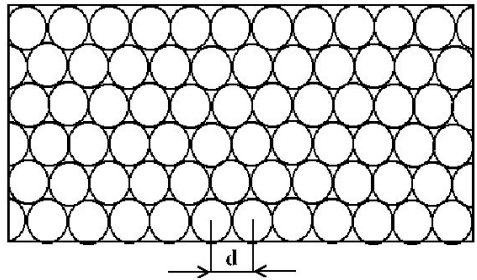

Fig.1. Common appearance of a globular photonic crystal, built of spherical particles (globules)

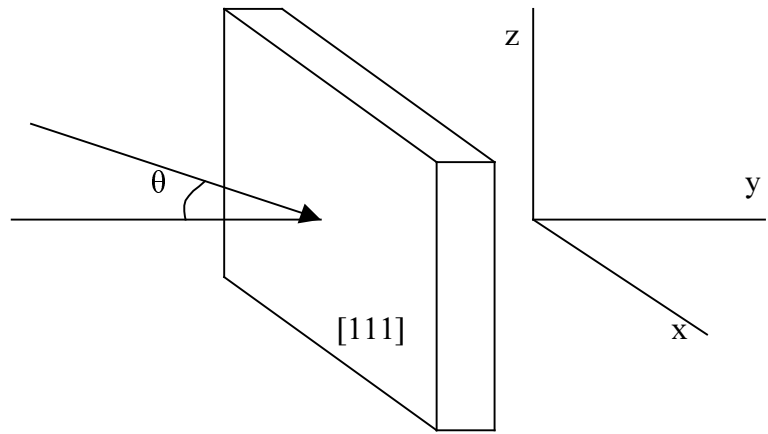

Fig.2. The scheme of illuminating the sample. Plane XY correspondes to the CU plate surface.

## 3.Observation of "photonic flame"

Opal crystals were placed on the Cu plate which was put into the cell with liquid nitrogen (see Fig.3). The number of crystals varied from 1 to 5. The distance (d) betwen the crystals was of the order of several centimeters (maximum value of d was 5 centimeters and was determined by the Cu plate size). One of the crystals was illuminated by the focused laser pulse. In the case of the reaching of the threshold visible (blue) luminescence appeared. The luminescence duration was from 1 to 4 seconds and it looked like inhomogeneous spot changing its spatial distribution and position on the surface of the crystal during this time.

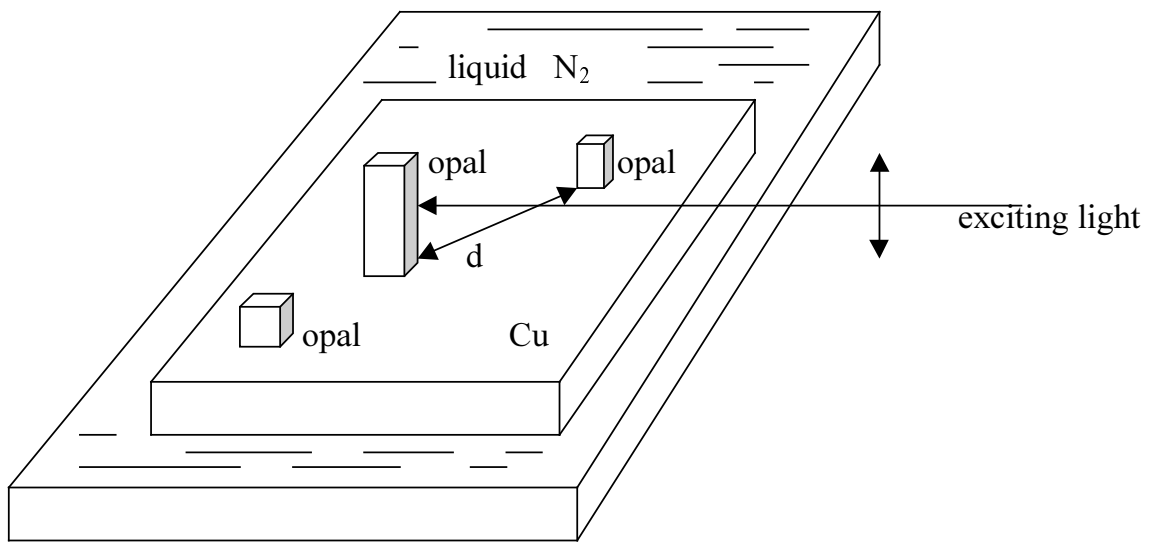

Fig.3. Experimental setup.

Parameters of the luminescence (duration, threshold) were determined by the geometric characteristic of the illumination and the refractive index contrast of the sample. For optimal geometry of the excitation the power density threshold for opal crystal was 0.12 Gw/cm$^2$, for opal crystals filled with ethanol – 0.05 Gw/cm², for opal crystal filled with acetone - 0,03 Gw/cm². Typical luminescence temporal distribution measured for the part of the crystal displaying the most intensive brightness is shown
on Fig.4. The same behavior is typical for all cases of the luminescence at these conditions of excitation, but the value of the luminescence duration fluctuated from shoot to shoot (~ 50%).

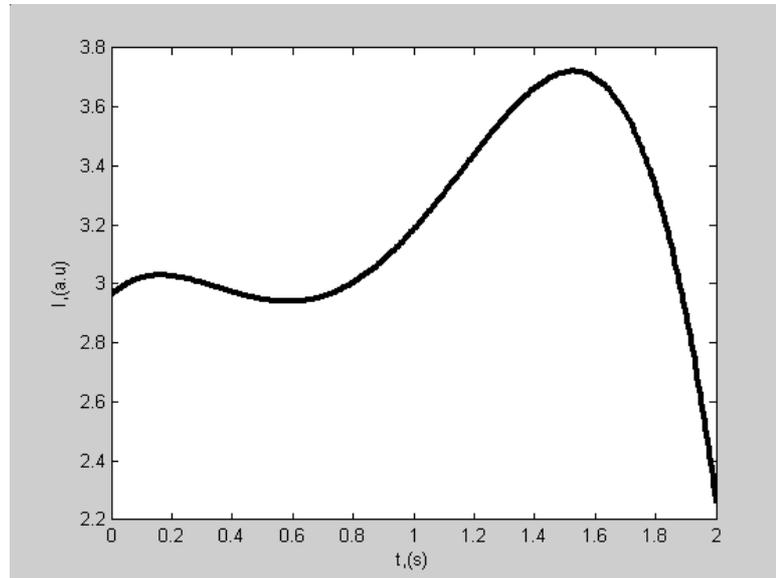

Fig.4. Temporal distribution of the visible luminiscence.

In the case of several crystals placed on the Cu plate only one of them was irradiated by the laser pulse. Luminescence took place in this crystal in the case of the threshold reaching. Bright shining of the other crystals began with some time delay after laser shoot. The value of this delay (and the intensity of the luminiscence) was determined by the spatial position of the crystals on the plate. The steal screen beeing put between the crystals (in order to avoid irradiating of the crystals by the light scattered by the crystal excited by the laser) did not stop the appearing of the luminescence if the distance between the Cu plate and the screen was more than 0.5 mm. The duration of the luminiscence was of the order of several seconds and temporal behavior was like shown on Fig.4. The typical features of such distribution were existence of maximum and large plato with near constant value of the intensity.

In order to show the role of the material of the plate used we repeated these measurements with plates of the same size but made from steel and quarz on which opal crystals were placed like in the previous experiments. Luminescence of the same kind in the irradiated crystal took place but the luminescence of the other samples situated on these plates was not observed.

The effect was also determined by the angle of incidence (Fig.2). For the samples used the value of the angle was chosen experimentally for achieving of the maximal value of the luminescence (it worth to mention that this value differed from 0 and was about $40^0$). Easier the effect was excited in the unprocessed samples. In Fig.5 one can see the luminescence of the crystals situated at the distance of about 1 centimeter from the crystal which was irradiated.

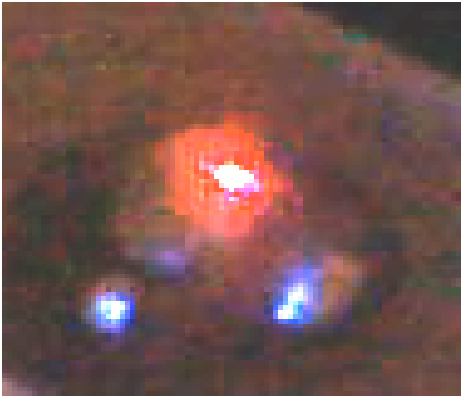 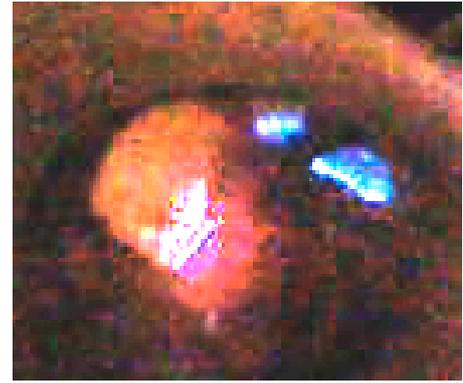

Fig.5 Visible luminiscence of the opal crystals in the case of the irradiating one of them (the irradiated crystal can be seen by bright red light; on the left picture it was the crystal in the center, on the right picture it was the crystal on the left). Left picture corresponds to the case where crystals are infiltrated by acetone. Right picture corresponds to the case of the opal crystals without infiltration.

## Conclusions.

In this paper we reported about novel effect – photonic flame effect. The main features of PFE are:
- At the excitation of the artificial opal crystal which is placed on the Cu plate at the temperature of the liquid nitrogen by the ruby laser pulse of the nanosecond range long-continued optical luminescence takes place in the case if the threshold of the process is reached;
- In the case of several opal crystals being put on the Cu plate while one of them is being irradiated bright visible luminescence occurs in all samples;
- Temporal behavior and thresholds of the luminescence have been determined. Photonic crystals infiltrated with different nonlinear liquids and without infiltration have been investigated. Investigated transport of the excitation between the samples spatially separated by the length of several centimeters gives the possibility of the practical applications of PFE.

The photonic flame effect can have different explanation. Probably an essential role is played by plasma properties. The slow transport of the excitations from the irradiated crystal to other photonic crystals can be associated with sound waves created due to laser pulse interaction with the sample. Exciton mechanism and surface waveguides on the surface of the Cu plate also can play important role. It was checked that the change of the properties of the plate surface was leading to change of the photonic flame effect. Removing the oxid layer from the plate changed the threshold PFE. Also some analogy with sonoluminiscence which inspired the recent studies of nonstationary Casimir effect (see , e.g.[17,18]) deserves a discussion. We will present the details of the photonic flame effect , e.g. spectral characteristics of the light emitted by the photonic crystals in future publication.

## Acknowledgments

This work was supported by the Byelorussian- Russian Foundation for Basic Research - BRFBR - Grant No 06-02-81024- Byel_a.